\documentclass[titlepage]{amsart}
\usepackage{graphicx}
\usepackage{amssymb}
\usepackage{comment}

\usepackage{epsfig} 
\usepackage{epstopdf}

 \usepackage{amsaddr}

\usepackage{amsmath} 
\usepackage{color}
\usepackage{subfig}

\usepackage{dblfloatfix}
\usepackage{fixltx2e}
\newenvironment{myequation}
{
    \begin{equation}
    \begin{aligned}
}
{
    \end{aligned}
    \end{equation}
}

\begin{document}

\markboth{Mohamed A. Khamis and Walid Gomaa and Basem Galal}{Deep learning is competing random forest in computational docking}

%
%

\title{Deep learning is competing random forest in computational docking}

\author{Mohamed A. Khamis}
\address{Cyber-Physical Systems Lab, \\
Egypt-Japan University of Science and Technology (E-JUST)\\
New Borg El-Arab City, Postal Code 21934 Alexandria, Egypt.}
\email{mohamed.khamis@ejust.edu.eg}

\author{Walid Gomaa}
\address{Cyber-Physical Systems Lab, \\
Egypt-Japan University of Science and Technology (E-JUST),\\
New Borg El-Arab City, Postal Code 21934 Alexandria, Egypt.}
\address{Faculty of Engineering, Alexandria University, Alexandria 21544, Egypt.}
\email{walid.gomaa@ejust.edu.eg}

\author{Basem Galal}
\address{Cyber-Physical Systems Lab, \\
Egypt-Japan University of Science and Technology (E-JUST)\\
New Borg El-Arab City, Postal Code 21934 Alexandria, Egypt.}
\email{mllover1992@gmail.com}

\maketitle


\begin{abstract}
Computational docking is the core process of computer-aided drug design; it aims at predicting the best orientation and conformation of a small molecule (drug ligand) when bound to a target large receptor molecule (protein) in order to form a stable complex molecule.
The docking quality is typically measured by a scoring function:
a mathematical predictive model that produces a score representing the binding free energy
and hence the stability of the resulting complex molecule.\\
We analyze the performance of both learning techniques on
the scoring power (binding affinity prediction),
the ranking power (relative ranking prediction),
docking power (identifying the native binding poses among computer-generated decoys), and
screening power (classifying true binders versus negative binders)
using the PDBbind 2013 database.\\
For the scoring and ranking powers, the proposed learning scoring functions depend on a wide range of features (energy terms, pharmacophore, intermolecular) that entirely characterize the protein-ligand complexes (about 108 features);
these features are extracted from several docking software available in the literature.
For the docking and screening powers, the proposed learning scoring functions depend on the intermolecular features of the RF-Score (36 features) to utilize a larger number of training complexes (relative to the large number of decoys in the test set).\\
For the scoring power, the DL\_RF scoring function (arithmetic mean between DL and RF scores) achieves Pearson's correlation coefficient between the predicted and experimentally measured binding affinities of 0.799 versus 0.758 of the RF scoring function.
For the ranking power, the DL scoring function ranks the ligands bound to fixed target protein with accuracy 54\% for the high-level ranking (correctly ranking the three ligands bound to the same target protein in a cluster) and with accuracy 78\% for the low-level ranking (correctly ranking the best ligand only in the cluster) while the RF scoring function achieves (46\% and 62\%) respectively. 
For the docking power, the DL\_RF scoring function has a success rate when the three best-scored ligand binding poses are considered within 2 \AA\ root-mean-square-deviation from the native pose of 36.0\% versus 30.2\% of the RF scoring function.
For the screening power, the DL scoring function has an average enrichment factor and success rate at the top 1\% level of (2.69 and 6.45\%) respectively versus (1.61 and 4.84\%) respectively of the RF scoring function.
\end{abstract}

keywords: Deep learning; Neural networks; Random forest; Drug discovery; Computational docking; Virtual screening.

\section{Introduction}




The core process of CADD is \emph{computational docking}.
Computational docking is the process of predicting the best orientation and conformation of a small molecule (drug ligand) when bound to a target large receptor molecule (protein) in order to form a stable complex molecule.
This amounts to predicting the \emph{binding free energy} (negative value in kcal/mol unit),
and hence the stability of the complex molecule resulting from the docking process.
A predicted \emph{binding affinity} inhibition constant $IC_{50}$, $K_i$, or $K_d$ (positive value in nanomolar unit) is then derived from the predicted binding free energy. This latter value is verified by comparing with the experimentally measured binding affinity.
The aim of the docking process is the activation/suppression of the target protein for/from preforming some \emph{functionality}.
The \emph{active site} (binding site) of the protein is the \emph{pocket} in which the atoms of the small ligand molecule (key) binds to the nearby amino-acids of the large protein molecule (lock).
In molecular docking, a large number of binding poses are evaluated using a \emph{scoring function}.
A \emph{scoring function} is a mathematical predictive model that produces a score that represents the \emph{binding free energy} of a binding pose.

In this paper, we show that a deep learning (DL) scoring function can compete one of the best learning scoring functions that is based on random forest (RF) with respect to
the scoring power (binding affinity prediction),
the ranking power (relative ranking prediction),
docking power (identifying the native binding poses among computer-generated decoys), and
screening power (classifying true binders versus negative binders)
on the well-known PDBbind benchmark \cite{WorldScientificIJPRAI:Wang2004PDBbindDatabase} version 2013.
Specifically, the contributions presented in this paper are the following:
(1) comparing our results with the best classical scoring functions presented in the latest comparative study (CASF-2013) published in mid-2014 \cite{WorldScientificIJPRAI:Li2014ComparativeAssessmentScoringFunctions}
whereas most of the similar published work compare with the previous CASF-2007;
the methods used for compiling the primary test set have been reformed resulting in an updated test set, i.e., the PDBbind core set (version 2013); this data set is the same size (195 protein-ligand complexes) as the one used in CASF-2007 but its content overlap with only 13\%,
(2) for the scoring and ranking powers, we use a wide range of features that entirely characterize the protein-ligand complexes; these features include
intermolecular features of RF-Score \cite{WorldScientificIJPRAI:Ballester2010PredictingProteinLigandBindingAffinity} (36 features),
energy terms of BALL software \cite{WorldScientificIJPRAI:Hildebrandt2010BALL} (5 features) and
energy terms of X-Score \cite{WorldScientificIJPRAI:Wang2002FurtherDevelopmentEmpiricalScoringFunctions} (8 features), and
pharmacophore features of SLIDE software \cite{WorldScientificIJPRAI:Zavodszky2002DistillingEssentialFeatures} (59 features), and
(3) for the docking and screening powers, we use the intermolecular features of the RF-Score \cite{WorldScientificIJPRAI:Ballester2010PredictingProteinLigandBindingAffinity} (36 features) to utilize a larger number of training complexes (relative to the large number of decoys in the test set) as well be discussed later.

The remainder of this paper is organized as follows.
In Section \ref{sec:Background}, we give the necessary chemical and molecular biology background with the state-of-the-art of relevant work.
Section \ref{sec:Methods} presents the training and testing data sets, the molecular features, the best classical scoring functions, the proposed learning scoring functions, and the evaluation methods.
Section \ref{sec:ResultsAndDiscussion} presents the evaluation results of the scoring, ranking, docking, and screening powers of the learning scoring functions.
Finally, Section \ref{sec:Conclusion} concludes the paper and proposes key points for research extensions.

\section{Background}
\label{sec:Background}



\subsection{Main capabilities of a scoring function}
The most important step in the docking process is scoring the \emph{conformations of a ligand}
in the corresponding \emph{binding site} of the receptor protein by using a \emph{scoring function}.
The binding affinity prediction using a \emph{scoring function} determines which binding mode is considered the best;
a \emph{scoring function} determines which ligand is considered the \emph{most effective drug}.

In \cite{WorldScientificIJPRAI:Brás2014ProteinLigandDockingInDrugDiscovery}, the authors present an introduction of the main available molecular docking methods with particular emphasis on the search algorithms and scoring functions.
In \cite{WorldScientificIJPRAI:Grinter2014ChallengesApplicationsRecentAdvancesProteinLigandDocking}, the authors introduce the protein-ligand docking methods used for structure-based drug design.
The authors discuss the fundamental challenges facing these methods and some of the current methodological topics of interest.

There are generally three main capabilities a reliable computational scoring function should satisfy \cite{WorldScientificIJPRAI:Ashtawy2012ComparativeAssessmentRankingAccuracies}:
(1) scoring power: the ability to produce scores for the different binding poses; these scores are supposed to be \emph{linearly correlated} with the experimentally measured binding affinities of the protein-ligand complexes of known 3D structures,
(2) ranking power: the ability to correctly \emph{rank} a given set of ligands with known binding poses when bound to a common protein, and
(3) docking power: the ability to identify the \emph{best binding pose} of a given ligand from a set of computationally generated poses when bound to a specific protein.
These three performance attributes were referred to by Cheng et al. \cite{WorldScientificIJPRAI:Cheng2009ComparativeAssessmentScoringFunctions}
as scoring power, ranking power, and docking power of a scoring function.
In \cite{WorldScientificIJPRAI:Li2014ComparativeAssessmentScoringFunctions}, the authors present also the \emph{screening power} which is the ability of a scoring function to identify the true binders to a given target protein among a pool of random molecules.

\subsection{Classical versus ML scoring functions}
Most scoring functions in use today can be categorized as either
\cite{WorldScientificIJPRAI:Ashtawy2012ComparativeAssessmentRankingAccuracies}:
(1) force field-based, (2) empirical-based, or (3) knowledge-based.
Force field scores are approximate molecular mechanics interaction energies, consisting of van der Waals and electrostatic components.
The parameters that define the intermolecular interactions are derived from experimental data and \emph{ab initio} simulations.
Empirical scoring functions adopt a different trend in calculating the binding free energy of the system.
The whole energy is assumed to be composed of weighted energy terms.
Linear regression methods are used to learn the coefficients of the model.
This can be done by fitting the known experimental binding energies to a \emph{training data set}.
Finally, a knowledge-based scoring function
is based on the theory that large databases of protein-ligand complexes can be statistically mined to deduce rules and models that are implicitly embedded in the data.

As mentioned in \cite{WorldScientificIJPRAI:Kinnings2011MachineLearningImproveDockingScoringFunction},
the physical-based and knowledge-based scoring functions are \emph{weak predictors} for the binding free energy
(and consequently for the binding affinity).
Traditional scoring functions assign a common set of weights to the individual energy terms
that contribute to the overall energy score.
However, as mentioned in \cite{WorldScientificIJPRAI:Kinnings2011MachineLearningImproveDockingScoringFunction},
the weights assigned to the individual energy terms that contribute to the \emph{overall energy score}
are in reality protein-family dependent.
Thus, in order to estimate a more accurate binding free energy (and consequently estimate the binding affinity),
a scoring function must be trained to derive a unique set of weights for each individual protein-family.
Moreover, traditional scoring functions improperly assume that the individual energy terms
contribute towards the total binding free energy in an additive manner.
Therefore, they predict the binding free energy from a \emph{linear combination} of the individual energy terms.
However, this is not theoretically sound
\cite{WorldScientificIJPRAI:Kinnings2011MachineLearningImproveDockingScoringFunction},
since it fails to consider the cooperative effects of the non-covalent interactions.
Thus, the scoring function have to model the non-linear relationships among the individual energy terms.

Machine learning is a paradigm shift that has proved itself in the context of virtual screening witnessed by the following.
First, improving the prediction ability of the binding affinity than traditional scoring functions (i.e., force field, empirical, etc.).
Second, the machine learning approach predicts the binding affinity based on some features of the protein-ligand complex molecule which are naturally available in the literature
(e.g., geometric features, physical force field energy terms, pharmacophore features, etc.).
In particular, the goal is to learn the relationship between these features
and the corresponding experimentally measured binding affinity given some training set of complex molecules.
Then, make use of this learned function to predict the binding affinity of new complexes
whose features are known but their experimental binding affinity is still unknown.
Recently, there are some non-parametric machine learning techniques used to model the \emph{functional form} of scoring functions given molecular databases, i.e., data-driven (not knowledge-based);
each complex structure is represented as a set of features that are relevant in predicting the complex binding affinity.

\subsection{Random forest scoring functions}
In \cite{WorldScientificIJPRAI:Ballester2010PredictingProteinLigandBindingAffinity}, the authors
use random forests \cite{WorldScientificIJPRAI:Breiman2001RandomForests} to learn how the \textit{atomic-level description} of the complex relates to the experimental binding affinity (RF-Score).
Each feature represents the number of occurrences of a particular protein-ligand atom type pair interacting within a certain distance range.
Four atom types were selected for proteins (C, N, O, S) and 9 atom types were selected for ligands (C, N, O, S, P, F, Cl, Br, I) considering all the common elemental atom types observed in protein-ligand complexes.
The authors in \cite{WorldScientificIJPRAI:Ballester2010PredictingProteinLigandBindingAffinity} achieve Pearson's correlation coefficient between the predicted and experimentally measured binding affinities of 0.774 on the PDBbind v2007 core set (N  = 195 complexes).

In \cite{WorldScientificIJPRAI:Zilian2013SFCscoreRF}, the authors present SFScoreRF that is an RF scoring function for improved affinity prediction of protein-ligand complexes. The authors in \cite{WorldScientificIJPRAI:Zilian2013SFCscoreRF} achieve Pearson's correlation coefficient between the predicted and experimentally measured binding affinities of 0.779 on the PDBbind v2007 core set (N  = 195 complexes).

In \cite{WorldScientificIJPRAI:Li2014istar}, the authors present a web platform for large-scale protein-ligand docking namely istar
that combined with RF-Score achieves a Pearson's correlation coefficient between the predicted and experimentally measured binding affinities of 0.855 on the PDBbind v2012 core set (N = 201 complexes).

In \cite{WorldScientificIJPRAI:Ballester2014MorePreciseChemicalDescription}, the authors investigate the impact of the chemical description of the complex on the predictive power of the resulting scoring function.
The authors in \cite{WorldScientificIJPRAI:Ballester2014MorePreciseChemicalDescription} achieve Pearson's correlation coefficient between the predicted and experimentally measured binding affinities of 0.803 on the PDBbind v2007 core set (N  = 195 complexes).

In \cite{WorldScientificIJPRAI:Li2014RFCyscore}, the authors show that replacing the linear regression used by Cyscore \cite{WorldScientificIJPRAI:Cao2014ImprovedProteinLigandBindingAffinityPrediction} by random forest (RF) can improve prediction performance.
In addition, the authors find that given sufficient training samples, RF comprehensively capture the non-linearity between structural features and measured binding affinities.
Moreover, the authors prove that incorporating more structural features and training with more samples can both boost RF performance.
The authors in \cite{WorldScientificIJPRAI:Li2014RFCyscore} use three sets of features:
Cyscore \cite{WorldScientificIJPRAI:Cao2014ImprovedProteinLigandBindingAffinityPrediction},
AutoDock Vina \cite{WorldScientificIJPRAI:Trott2010AutoDockVina}, and
RF-Score \cite{WorldScientificIJPRAI:Ballester2010PredictingProteinLigandBindingAffinity}.
Cyscore compromises four numerical features: $\Delta G_\text{hydrophobic}$, $\Delta G_\text{vdw}$, $\Delta G_\text{hbond}$ and $\Delta G_\text{entropy}$.
AutoDock Vina compromises six numerical features: $Gauss_1$, $Gauss_2$, $Repulsion$, $Hydrophobic$, $HBonding$ and $N_\text{rot}$.
RF-Score compromises 36 features defined as the occurrence of intermolecular contacts between two elemental atom types.
The authors in \cite{WorldScientificIJPRAI:Li2014RFCyscore} achieve Pearson's correlation coefficient between the predicted and experimentally measured binding affinities of 0.803 on the PDBbind v2007 core set (N  = 195 complexes).

In \cite{WorldScientificIJPRAI:Wang2014ComparativeStudyFamilySpecificProteinLigandAffinityPrediction}, the authors proposed RF scoring function to predict the protein-ligand binding affinity based on a comprehensive feature set covering protein sequence, binding pocket, ligand structure, and intermolecular interaction.
Feature processing and reduction are performed for different protein family datasets. 
Three family-specific models were constructed for three important protein target families of HIV-1 protease, trypsin and carbonic anhydrase respectively.
For comparison, two generic models including diverse protein families were also built.
The evaluation results show that models on family-specific datasets have the superior performance to those on the generic datasets and the Pearson's correlation coefficient on the test sets are 0.740, 0.874, 0.735 for HIV-1 protease, trypsin and carbonic anhydrase respectively.

In \cite{WorldScientificIJPRAI:Ashtawy2012ComparativeAssessmentRankingAccuracies}, the authors assess the ranking accuracy of ML scoring functions and classical scoring functions using both 2007 and 2010 PDBbind benchmark data sets \cite{WorldScientificIJPRAI:Wang2004PDBbindDatabase};
working on both diverse and protein-family specific test sets.
The best ML scoring function (based on RF) ranks the ligands correctly based on their experimentally measured binding affinities with accuracy 62.5\% and identifies the top binding ligand with accuracy 78.1\%.
For each protein-ligand complex, the authors in \cite{WorldScientificIJPRAI:Ashtawy2012ComparativeAssessmentRankingAccuracies} extracted features using:
X-Score (6 features) \cite{WorldScientificIJPRAI:Wang2002FurtherDevelopmentEmpiricalScoringFunctions},
AffiScore (30 features) \cite{WorldScientificIJPRAI:Zavodszky2002DistillingEssentialFeatures, WorldScientificIJPRAI:Schnecke2002VirtualScreeningSolvation, WorldScientificIJPRAI:Zavodszky2005SideChainFlexibility},
and RF-Score (36 features) \cite{WorldScientificIJPRAI:Ballester2010PredictingProteinLigandBindingAffinity}.
The authors exploit six ML scoring functions:
multiple linear regression (MLR),
multivariate adaptive regression splines (MARS) \cite{WorldScientificIJPRAI:Trevor2010MultivariateAdaptiveRegressionSplineModels},
$k$-nearest neighbors (kNN) \cite{WorldScientificIJPRAI:Schliep2010WeightedK-NearestNeighbors},
support vector machines (SVM) \cite{WorldScientificIJPRAI:Dimitriadou2010MiscellaneousFunctions},
random forests (RF) \cite{WorldScientificIJPRAI:Breiman2001RandomForests}, and
boosted regression trees (BRT) \cite{WorldScientificIJPRAI:GeneralizedBoostedRegressionModels}.
The results of the experiments in \cite{WorldScientificIJPRAI:Ashtawy2012ComparativeAssessmentRankingAccuracies}
conclude that utilizing as many relevant features as possible in conjunction with ensemble-based approaches
like BRT and RF (which are resilient to over-fitting) is the best option.

In \cite{WorldScientificIJPRAI:Khamis2015ComparativeAssessmentMLScoringFunctionsOnPDBbind2013}, we present a comparative assessment of scoring, ranking, docking, and screening powers of ML scoring functions on the PDBbind v2013 and compared with the classical scoring functions presented in \cite{WorldScientificIJPRAI:Li2014ComparativeAssessmentScoringFunctions}.
For a more comprehensive survey about the use of the most prominent ML techniques in computational docking, the readers are referred to \cite{WorldScientificIJPRAI:Khamis2015MachineLearningInComputationalDocking}.

\subsection{Beware of RF scoring functions}
Despite the superiority of RF scoring functions to predict the experimental binding constants from protein-ligand X-ray structures of the PDBbind dataset, the ranking, docking, and screening powers of RF scoring functions should also be examined.
In \cite{WorldScientificIJPRAI:Gabel2014BewareMachineLearningScoringFunctions},
the authors present RF scoring function trained on simple descriptors that outperforms a prototype scoring function in predicting the binding constants from the atomic coordinates (scoring power test);
the authors in \cite{WorldScientificIJPRAI:Gabel2014BewareMachineLearningScoringFunctions} achieve Pearson's correlation coefficient between the predicted and experimentally measured binding affinities of 0.791 on the PDBbind v2007 core set (N = 195 complexes).
However, the proposed RF scoring function does not discriminate DUD-E \cite{WorldScientificIJPRAI:Mysinger2012DUDE} actives from decoys in docking experiments (virtual screening power test).
Moreover, the proposed RF scoring function is insensitive to docking pose accuracy (docking power test).

In addition, the work presented in \cite{WorldScientificIJPRAI:Ashtawy2015BgNScoreBsNScore} proves that ensemble NN scoring functions are more accurate in predicting the binding affinity of protein-ligand complexes than RF.
Particularly, neural networks have the ability to approximate any underlying function smoothly in contrast to decision trees that model functions with step changes across decision boundaries.
Thus, we propose a deep learning scoring function combined with dropout that can be considered a model averaging technique over a large number of neural networks that outperforms RF scoring function.
Hence, the next section presents a comprehensive background on deep learning and its superiority in other application domains.

\subsection{Deep Learning}
Our work is based on recent successful algorithms in unsupervised feature learning and deep learning    \cite{WorldScientificIJPRAI:Hinton2006FastLearningAlgorithm,WorldScientificIJPRAI:Bengio2007GreedyLayerWiseTraining}.
Deep learning is a paradigm that focuses on learning multiple levels of representation and abstraction that help making sense of data such as images, sound, and text. Deep learning systems achieved state-of-the-art performance on numerous machine learning tasks specially in image and speech recognition.

One of the largest neural networks are built by connecting 16,000 cores \cite{WorldScientificIJPRAI:QuocV2013BuildingHighLevelFeatures} to learn multiple levels of representations of images from a large data-set (10 million images selected randomly from the internet).
The training was completely unsupervised. The resulting neurons function as detectors for faces, human body, and cat faces. These results proved that computers can learn to detect faces using only unlabeled data.

Recursive neural networks also outperforms the state-of-the-art approaches in segmentation, annotation, and scene classification \cite{WorldScientificIJPRAI:AndrewNg2011ParsingNaturalScenesAndNaturalLanguage}.
The recurrent neural networks are used to build a new system called DeepSpeech \cite{WorldScientificIJPRAI:AndrewNg2014DeepSpeech} that outperformed the previously published results on the widely studied Switchboard Hub5'00 and commercial speech systems.

Deep learning methods are new to the biological field.
Recently, deep learning methods are applied to residue-residue contact prediction and disorder prediction \cite{WorldScientificIJPRAI:Eickholt2012PredictingProteinResidueContacts,WorldScientificIJPRAI:Eickholt2013DNdisorder,WorldScientificIJPRAI:DiLena2012DeepArchitecturesProteinContactMapPrediction}.
In addition,  a couple of deep learning protein structure predictors have been developed, including a multifaceted prediction tool that predicts several protein structural elements in tandem \cite{WorldScientificIJPRAI:Qi2012UnifiedMultitaskArchitecture}.

\section{Methods}
\label{sec:Methods}
\subsection{Training and testing data sets}
The PDBbind benchmark \cite{WorldScientificIJPRAI:Wang2004PDBbindDatabase} could be considered the most widely used for binding affinity prediction \cite{WorldScientificIJPRAI:Li2014RFCyscore}.
Since its first public release in 2004, the PDBbind database has been updated on an annual basis.
For instance, PDBbind v2013 provides experimental binding affinity data for 10776 biomolecular complexes in PDB,
including 8302 protein-ligand complexes and 2474 other types of complexes.
In \cite{WorldScientificIJPRAI:Liu2015PDBwideCollectionBindingData}, the authors describe the current methods used for compiling PDBbind.
The authors also review some typical applications of PDBbind published in scientific literature.
For each protein-ligand complex, the experimentally measured binding affinity either the dissociation constant $K_d$ or inhibition constant $K_i$ was manually collected from its primary literature reference \footnote{The experimentally measured binding affinity either the dissociation constant $K_d$ or inhibition constant $K_i$ used in the present article are extracted from the file named pdbbind\_2013\_refined.pdf that was initially posted on the PDBbind-CN web site (http://www.pdbbind-cn.org/).}.

For detailed information on the compilation of the PDBbind benchmark (version 2013), the reader is referred to \cite{WorldScientificIJPRAI:Li2014ComparativeAssessmentScoringFunctionsCompilation}.
Here we mention briefly the main rules for selecting qualified protein-ligand complexes into the PDBbind refined set (version 2013) \cite{WorldScientificIJPRAI:Li2014ComparativeAssessmentScoringFunctionsCompilation}:
(1) only complexes with crystal structures are accepted,
(2) resolution of the complex structure must be better than 2.5 \AA,
(3) if any fragment on the ligand molecule is missing in the crystal structure, the complex is not included,
(4) if any side chain fragment is missing at the protein binding site within 8 \AA\ from the ligand, the complex is not included,
(5) complexes with known dissociation constants ($K_d$) or inhibition constant ($K_i$) are accepted, while complexes with only half-inhibition or half-effect concentrations ($IC_{50}$ or $EC_{50}$) values are not included,
(6) the binding site on the protein molecule must not contain any non-standard amino acid residues in direct contact with the bound ligand in distance $<$ 5 \AA, and
(7) if the buried surface area of the ligand molecule is below 15\% of its total surface area, the complex is not included.

The PDBbind refined set (version 2013) was selected out of 8302 protein-ligand complexes recorded in the PDBbind general set (version 2013) through a complicated process \cite{WorldScientificIJPRAI:Li2014ComparativeAssessmentScoringFunctionsCompilation} resulting in 2959 protein-ligand complexes.
Finally, qualified complexes are clustered by 90\% similarity in protein sequences.
The binding constants of these complexes $\log K_a$ span nearly 10 orders of magnitude ($\log K_a = 2.07 \sim 11.52$).
The resulting 195 complexes are grouped into 65 clusters by protein sequences, namely PDBbind core set (version 2013).
Each cluster consists of three complexes, which are referred to as ``the best'', ``the median'', and ``the poorest'' by their binding affinities. The binding affinity of the best complex is required to be at least 100 times higher than that of the poorest.
Because of the structural diversity of the core set, it is a common practice to use the core set as a test set and the remaining complexes in the refined set as a training set \cite{WorldScientificIJPRAI:Li2014RFCyscore}.

Decoy ligand binding poses were needed in CASF-2013 \cite{WorldScientificIJPRAI:Li2014ComparativeAssessmentScoringFunctions} to evaluate the docking and screening powers of each scoring function.
In CASF-2013, the authors used three popular molecular docking programs, including GOLD (version 5.1, Cambridge Crystallographic Data Center), Surflex implemented in the SYBYL software (version 8.1, CERTARA Inc.) and the molecular docking module implemented in the MOE software (version 2011, Chemical Computing Group).
These 3 docking programs use different algorithms for sampling ligand binding poses.
The key parameters and settings used in docking are described in detail in \cite{WorldScientificIJPRAI:Li2014ComparativeAssessmentScoringFunctions};
these parameters were chosen to generate diverse rather than converged binding poses.

The outcomes of all three docking programs were combined to obtain a set of ligand binding poses for each complex.
Only the binding poses with root-mean-square deviation (RMSD) values lower than 10 \AA\ (in respect to the native binding pose) were considered at subsequent steps.
All of these binding poses were grouped by their RMSD values (0 $\sim$ 10 \AA) into 10 bins with an interval of 1 \AA.
The binding poses in each bin were further grouped into up to 10 clusters according to their initial similarities.
In each cluster, the binding pose with the lowest internal strain energy was selected as the representative of that cluster.
Through the above process, ideally a total of 100 representative decoy ligand poses would be obtained for each complex.
However, the number of final selected decoy binding poses was actually lower than 100 for many protein-ligand complexes because of the geometrical constraints of the binding site or the parametric shape of the ligand molecule.

In \cite{WorldScientificIJPRAI:Khamis2015ComparativeAssessmentMLScoringFunctionsOnPDBbind2013}, we used two commands available in the BALL molecular software \cite{WorldScientificIJPRAI:Hildebrandt2010BALL} for decoys pre-processing:
(1) \emph{LigandFileSplitter} command that splits the decoys.mol2 file of each complex one molecule per file, and
(2) \emph{Converter} command that converts each individual mol2 file to sdf format.
We then customized the RF-Score feature extraction code initially available by \cite{WorldScientificIJPRAI:Ballester2010PredictingProteinLigandBindingAffinity}
for taking as an input the protein PDB file and the decoy sdf file
and generating the corresponding 36 intermolecular features for every protein-ligand pair.

\subsection{Molecular features}
For the scoring and ranking powers, the proposed ML scoring functions depend on wide range of features that entirely characterize the protein-ligand complexes.
These features include
intermolecular features of the RF-Score \cite{WorldScientificIJPRAI:Ballester2010PredictingProteinLigandBindingAffinity} (36 features),
energy terms of the BALL software \cite{WorldScientificIJPRAI:Hildebrandt2010BALL} (5 features) and
energy terms of the X-Score \cite{WorldScientificIJPRAI:Wang2002FurtherDevelopmentEmpiricalScoringFunctions} (8 features), and
pharmacophore features of the SLIDE software \cite{WorldScientificIJPRAI:Zavodszky2002DistillingEssentialFeatures} (59 features).
The RF-Score \cite{WorldScientificIJPRAI:Ballester2010PredictingProteinLigandBindingAffinity} results in 36 intermolecular features;
the occurrence count of intermolecular contacts between two elemental atom types in every protein-ligand pair.
For proteins 4 atom types (C, N, O, S) and for ligands 9 atom types (C, N, O, S, P, F, Cl, Br, I) were selected to generate features considering all the common elemental atom types observed in protein-ligand complexes.

The BALL software \cite{WorldScientificIJPRAI:Hildebrandt2010BALL} results in 5 energy terms;
advanced electrostatics, van der Waals, fragmentational solvation, hydrogen bond, and (nRot$>$14).
Two BALL commands are needed to obtain those features;
\emph{WaterFinder}\footnote{The protein resulting from the WaterFinder tool is used as input for the X-Score and SLIDE software for consistency. Also, since water molecules are often important for binding of ligands, it is advisable to use WaterFinder.} finds the strongly bound waters and
\emph{ConstraintsFinder} finds the strongly interacting residues.

The X-Score software \cite{WorldScientificIJPRAI:Wang2002FurtherDevelopmentEmpiricalScoringFunctions} results in 8 energy terms;
\begin{enumerate}
    \item ligand molecular weight,
    \item van der Waals,
    \item hydrogen bonding,
    \item hydrophobic pairwise contacts,
    \item hydrophobic ligand atoms match inside hydrophobic binding site,
    \item hydrophobic surface area of ligand buried upon binding,
    \item contribution of the number of rotors for the ligand, and
    \item octanol-water partition coefficients (log P).
\end{enumerate}
This includes preparing the input PDB file using the \emph{fixpdb} tag of the \emph{xscore} command
then preparing the input mol2 file using the \emph{fixmol2} tag,
then scoring the protein-ligand complex using the \emph{score} tag,
and finally calculate the logp value of the ligand using the \emph{logp} tag.

The SLIDE software \cite{WorldScientificIJPRAI:Zavodszky2002DistillingEssentialFeatures} results in 59 features of pharmacophore nature;
\begin{enumerate}
    \item ligand hydrophobic cluster points count,
    \item ligand acceptor cluster points count,
    \item ligand donor cluster points count,
    \item ligand doneptor cluster points count,
    \item protein hydrophobic surface points count,
    \item protein hydrophobic cluster points count,
    \item protein acceptor cluster points count,
    \item protein donor cluster points count,
    \item protein doneptor cluster points count,
    \item protein unbumped hydrophobic cluster points count,
    \item protein metal acceptor points count,
    \item protein acceptor cluster points remain count,
    \item protein donor cluster points remain count,
    \item protein doneptor cluster points remain count,
    \item protein hydrophobic cluster points remain count,
    \item protein-ligand hydrophobic contacts,
    \item protein-ligand H-bond count,
    \item protein-ligand salt-bridge count,
    \item number of ligand neighbors,
    \item number of ligand carbons,
    \item number of exposed ligand carbons,
    \item remaining Van der Waals collisions,
    \item total Van der Waals overlap (\AA),
    \item anchor fragment translations,
    \item side-chain mean-field iterations,
    \item ligand side-chain rotations,
    \item protein side-chain rotations,
    \item number of ligand non-hydrogen atoms,
    \item total sum of hydrophobicity values for all hydrophobic protein-ligand atom pairs,
    \item total difference in hydrophobicity values between all hydrophobic protein-ligand atom pairs,
    \item total difference in hydrophobicity values between all protein-ligand hydrophobic/hydrophilic mismatches,
    \item total sum of hydrophobicity values for all hydrophilic protein-ligand atom pairs,
    \item total difference in hydrophobicity values between all hydrophilic protein-ligand atom pairs,
    \item number of interfacial ligand atoms,
    \item number of exposed hydro ligand atoms,
    \item number of ligand flexible bonds,
    \item number of all interfacial flexible bonds,
    \item total average hydrophobicity values for all ligand atoms relative to all neighboring hydrophobic protein atoms,
    \item sum of hydrophobicity values for all hydrophobic protein-ligand atom pairs (in old Slide version),
    \item number of flexible interfacial ligand bonds,
    \item sum of the degree of similarity between protein-ligand hydrophobic atom contacts,
    \item pairwise contact of hydrophilic-hydrophilic protein-ligand contacts,
    \item pairwise contact of hydrophobic-hydrophilic protein-ligand contacts,
    \item total of protein hydrophobicity values for protein atoms involved in hydrophobic-hydrophobic contacts,
    \item sum of hydrophobic atom hydrophobicity values for hydrophobic/hydrophilic mismatch pairs,
    \item total hydrophilicity of all protein interfacial atoms,
    \item distance normalized version of protein hydrophobicity values for protein atoms involved in hydrophobic-hydrophobic contacts,
    \item total of protein hydrophobicity values for protein atoms at the interface,
    \item distance normalized version of protein hydrophobicity values for protein atoms at the interface,
    \item total of ligand hydrophilicity values for interfacial ligand atoms,
    \item number of hydrophobic-hydrophobic contacts,
    \item number of hydrophobic-hydrophilic contacts,
    \item increase in hydrophobic environ,
    \item number of intra target salt bridges,
    \item number of ligand uncharged polar atoms at the protein-ligand interface which do not have a bonding partner,
    \item number of protein uncharged polar atoms at the protein-ligand interface which do not have a bonding partner,
    \item number of ligand charged atoms at the protein-ligand interface which do not have a bonding partner,
    \item number of protein charged atoms at the protein-ligand interface which do not have a bonding partner, and
    \item total buried hydrophobic contacts of the ligand upon binding.
\end{enumerate}

This includes dehydrogenating the input pdb file using the \emph{pdbdehydrogen} command
then setup the database for biased (ligand-based) template using the \emph{biased} tag of the \emph{setup\_dbase} command,
then generating the biased (ligand-based) template using the \emph{biased} tag of the \emph{temp\_gen} command,
then setup the database for unbiased (protein-based) template using the \emph{unbiased} tag of the \emph{slide\_setup.pl} command,
then generate the unbiased (protein-based) template using the \emph{unbiased} tag of the \emph{temp\_gen} command.


A biased template represents known ligand interactions, and an unbiased template represents additional opportunities for making good interactions in the protein binding site. Thus, we merged the two templates into one for screening databases.
Firstly, we set up the directory to hold the new merged template using the \emph{merged} tag of the \emph{slide\_setup.pl} command. Secondly, we concatenated both of the unbiased and biased templates using the \emph{cat} command.
Finally, we run slide based on the merged template using the \emph{merged} tag of the \emph{run\_slide} command.

RF-Score was able to retrieve the 36 features on the entire training set 2764 complexes (2959-195) and the entire test set 195 complexes.
BALL was able to retrieve the 5 features on 2481 complexes of the training set and 181 complexes of the test set.
X-Score was able to retrieve the 8 features on 2507 complexes of the training set and 180 complexes of the test set.
SLIDE was able to retrieve the 59 features on 2301 complexes of the training set and 165 complexes of the test set.
The final number of common protein-ligand complexes resulting from the four sources of features (RF-Score, BALL, X-Score, and SLIDE) was 2281 complexes in the training set and 164 complexes in the test set.

For the docking and screening powers, the proposed ML scoring functions depend on the intermolecular features of the RF-Score \cite{WorldScientificIJPRAI:Ballester2010PredictingProteinLigandBindingAffinity} (36 features) for mainly two reasons:
(1) utilizing a larger number of training complexes (2764 complexes) relative to the large number of decoys in the test set (docking test set = 15,821 complexes and screening test set = 63,5729 complexes),
(2) comparing the presented results with the recently published work in \cite{WorldScientificIJPRAI:Gabel2014BewareMachineLearningScoringFunctions} which depends mainly on RF-Score features \cite{WorldScientificIJPRAI:Ballester2010PredictingProteinLigandBindingAffinity} with several modifications.

\subsection{Best classical scoring functions}
As mentioned in the introduction section, we compared the performance of the proposed learning scoring functions with the best classical scoring functions presented in the latest comparative study (CASF-2013) published in mid 2014 \cite{WorldScientificIJPRAI:Li2014ComparativeAssessmentScoringFunctions}.
The authors in \cite{WorldScientificIJPRAI:Li2014ComparativeAssessmentScoringFunctions} present a panel of 20 scoring functions tested on PDBbind v2013 benchmark.
In this panel, 18 scoring functions are implemented in commercial molecular modeling software.
In addition, X-Score (version 1.3) is an academic scoring function \cite{WorldScientificIJPRAI:Wang2002FurtherDevelopmentEmpiricalScoringFunctions} where the coefficients before each energy term are re-calibrated by \cite{WorldScientificIJPRAI:Li2014ComparativeAssessmentScoringFunctions} on the remaining 2,764 protein-ligand complexes in the PDBbind refined set (version 2013) after removing the 195 protein-ligand complex in the PDBbind core set.
In addition, there are three optional modes in X-Score (version 1.3), i.e., $\text{X-Score}^\text{HM}$, $\text{X-Score}^\text{HP}$, and $\text{X-Score}^\text{HS}$, which differ from each other only in the hydrophobic effect term.
The performance of $\text{X-Score}^\text{HM}$ was slightly better than the other two modes in most tests in CASF-2013.
Thus, $\text{X-Score}^\text{HM}$ was chosen to represent X-Score in CASF-2013.
Besides, CASF-2013 introduced a naive scoring function as a reference that uses a single descriptor, i.e., buried solvent-accessible surface area of the ligand molecule upon binding ($\Delta$ SAS) which is an estimation of the size of protein-ligand binding interface.

For the scoring power test, $\text{X-Score}^\text{HM}$ was the best classical scoring function where the Pearson's correlation coefficient between the predicted and experimentally measured binding affinities was 0.614.
For the ranking power, $\text{X-Score}^\text{HM}$ ranks the ligands bound to fixed target protein with accuracy 58.5\% for the high-level ranking (correctly ranking the three ligands bound to the same target protein in a cluster) and with accuracy 72.3\% for the low-level ranking (correctly ranking the best ligand only in the cluster). For fair comparison between the best classical scoring function $\text{X-Score}^\text{HM}$ and our proposed learning scoring functions, we re-generated the previous results on the 164 complexes for which we succeeded to generate features \cite{WorldScientificIJPRAI:Khamis2015ComparativeAssessmentMLScoringFunctionsOnPDBbind2013}. The following are the updated results of $\text{X-Score}^\text{HM}$; Pearson's correlation coefficient = 0.611, high-level ranking = 54\%, low-level ranking = 70\%.

For the docking power, the best classical scoring function ChemPLP@GOLD has a success rate in identifying the top best-scored ligand binding pose within 2 \AA\ RMSD from the native pose of 81.0\%.
For the screening power, the best classical scoring function GlideScore-SP has an average enrichment factor and success rate at the top 1\% level of 19.54 and 60\% respectively.

\subsection{Machine learning methods: deep learning \& random forest}
We analyze the performance of the learning scoring functions with respect to their scoring, ranking, docking, and screening powers.
We built a sparse deep auto-encoder to learn multiple layers of representation; the input layer is the 108 features (intermolecular, energy terms, pharmacophore) that characterize the protein-ligand complexes, then we used greedy layer-wise training algorithm to pre-train our model.
In the greedy layer-wise training, each layer is pre-trained to produce a higher level representation of the raw input based on the received representation from the previous layer, which is then fine-tuned using gradient descent.

For the DL scoring function, we used the \emph{Deep learning toolbox} released by Rasmus Berg Palm \cite{WorldScientificIJPRAI:Palm2012LearningDeepHierarchicalModels} that implements models for deep learning, e.g., stacked auto-encoders, deep belief network, convolutional neural nets, and various pre-processing functions.
We used the dropout technique \cite{WorldScientificIJPRAI:Srivastava2014Dropout} in order to reduce overfitting. The key idea of this technique is to randomly drop units in training with probability $p$, then sample the final model from exponential number of thinned networks.
At test time, we use single neural network without dropout but each outgoing weights of the units are multiplied by $p$.
This technique improves the performance of our neural network and outperforms other regularization methods.

For the scoring and ranking powers, we trained stacked auto-encoders (SAE) neural network of size [$108 \times 500 \times 500$] with hyperbolic activation hidden units on unlabeled training data-set of size [$2280 \times 108$] to learn hierarchical features using greedy layer-wise training method with learning rate = 0.01, and dropout rate = 0.8.
Then, we used the trained SAE to initialize neural network of size [$108 \times 500 \times 500 \times 1$] with linear output layer. Finally, we fine-tuned the neural network using mini-batch gradient descent with batch size = 228, learning rate = 0.002, momentum = 0.9, and dropout rate in fine-tuning = 0.9; this neural network converges after 670 epochs.
For the docking and screening powers, the parameter optimization was done on 4 parameters:
learning rate = 0.003, momentum = 0.9, L2 weight decay = 0.005, and dropout rate = 0.9;
other parameters are: using 600 epochs, architecture size = [$35 \times 400 \times 400 \times 1$], and batch size = 691.

For the RF scoring function, we used the \emph{randomForest} function in the \emph{randomForest} package \cite{WorldScientificIJPRAI:Liaw2002ClassificationRegressionRandomForest} and the RF-Score code available by \cite{WorldScientificIJPRAI:Ballester2010PredictingProteinLigandBindingAffinity}.
In \emph{bagging}, we take a subset of training data (select randomly \emph{n} training data out of \emph{N} training data with replacement at each decision node of the tree) and select randomly \emph{m} input features out of \emph{M} input features) to train up each tree; after multiple trees are trained, a voting scheme is used to predict testing data;
random forest (RF) is one of the most popular \emph{bagging} methods \cite{WorldScientificIJPRAI:BigDataMachineLearning}.
The tuning parameters of RF include:
the $n_\text{tree}$ that is the number of trees to grow (chosen by 500), and
the $m_\text{try}$ that is the number of variables randomly sampled as candidates at each split;
$m_\text{try}$ $\in$ \{2: number of components\}.
For the scoring and ranking powers: $m_\text{best}$ = 96.
For the docking and screening powers: $m_\text{best}$ = 7.

For fair comparison between the learning scoring functions, we used the same training set (one complex was excluded at random to make the training set size divisible to equally-sized batches, i.e., the training set size equals DL batch size $\times$ 10). Also, we used the same training complexes shuffling technique for the learning methods in data preparation.

\subsection{Evaluation methods of scoring power}
As mentioned in CASF-2013, the scoring power refers to the ability of a scoring function to produce binding scores in a linear correlation with the experimentally measured binding affinities.
In CASF-2013, this feature was evaluated on the known 3D structures of the 195 protein-ligand complexes in the test set including the original crystal structures and locally optimized complex structures.
However, since it is mentioned in CASF-2013 that the performance of most scoring functions retains at basically the same level or gets slightly worse on the optimized complex structures, we found that it is sufficient to compare our results with those of the crystal structures only (for both of the scoring and ranking powers).
In CASF-2013, the scoring power of a scoring function was evaluated by the classic Pearson's correlation coefficient (R) between its binding score and the experimental binding affinities (Eq. \ref{eq:PearsonCorrelationCoefficient}) and the standard deviation (SD) in regression (Eq. \ref{eq:StandardDeviation}) \cite{WorldScientificIJPRAI:Li2014ComparativeAssessmentScoringFunctions}:

\begin{myequation}
 R = \frac{\sum {(x_i - \bar{x}) (y_i - \bar{y})}}{\sqrt{\sum(x_i - \bar{x})^2} \sqrt{\sum(y_i - \bar{y})^2}}
\label{eq:PearsonCorrelationCoefficient}
\end{myequation}

\begin{myequation}
 SD = \sqrt{\frac{\sum {[y_i - (a + b . x_i)]^2}}{N - 1}}
\label{eq:StandardDeviation}
\end{myequation}

where, $x_i$ is the binding score computed by a certain scoring function on the $i^{th}$ complex;
$y_i$ is the experimental binding affinity of this complex;
$a$ and $b$ are the intercept and the slope of the regression line between the computed binding score and the experimental one respectively; the binding constants are all given in the logarithm units ($\log K_a$) \cite{WorldScientificIJPRAI:Li2014ComparativeAssessmentScoringFunctions}.

\subsection{Evaluation methods of ranking power}
As mentioned in CASF-2013, the ranking power refers to the ability of a scoring function to correctly rank the known ligands relative to the same target protein by their binding affinities; 
the test set used in CASF-2013 consists of 65 clusters of complexes, each of which has 3 complexes formed by the same target protein where the binding affinity of the best complex is required to be at least 100 times higher than that of the poorest.

As mentioned in \cite{WorldScientificIJPRAI:Ashtawy2012ComparativeAssessmentRankingAccuracies}, for test sets that are composed of one protein family (one cluster) bound to a diverse set of ligands, one can interpret the Spearman rank correlation coefficients ($\text{R}_\text{s}$) as an alternative measure for ranking accuracy; this measure represents the correlation between ranks based on the predicted binding affinities and the ranks based on the experimentally measured binding affinities; i.e.,
a scoring function that achieves higher $\text{R}_\text{s}$ value for some data set is considered more accurate (in terms of ranking power) than its counterparts having smaller $\text{R}_\text{s}$ values.

As mentioned in CASF-2013, if a scoring function correctly ranked the three complexes in a specific cluster as ``the best $>$ the median $>$ the poorest'', one point is recorded for this scoring function, i.e., ``high-level'' ranking;
an overall success rate is computed accordingly over the entire test set \cite{WorldScientificIJPRAI:Li2014ComparativeAssessmentScoringFunctions}.
In order to provide an additional index, a ``low-level'' success rate is considered when a scoring function is able to only rank the best complex as the top one in the cluster regardless of the ranking of the median and the poorest complexes in the same cluster \cite{WorldScientificIJPRAI:Li2014ComparativeAssessmentScoringFunctions}.

\subsection{Evaluation methods of docking power}
As mentioned in CASF-2013, the docking power refers to the ability of a scoring function to identify the native binding pose among computer-generated decoys;
ideally, the native binding pose is identified as the one with the best binding score;
a set of decoy binding poses (up to 100) was generated for each protein-ligand complex in the PDBbind v2013 core set (195 complexes) by using several molecular docking programs.
Each scoring function (in a panel of 20 scoring functions \cite{WorldScientificIJPRAI:Li2014ComparativeAssessmentScoringFunctions}) was applied to score the decoy set of each protein-ligand complex and finds consequently the best-scored binding pose.
The decoy set of each complex includes the native binding pose to ensure that there exists at least one correct binding pose.
If the RMSD value between the native binding pose and the best-scored binding pose among all decoys plus the native one (computed using Eq. \ref{eq:RMSD}) fell below predefined cutoff, e.g., RMSD $<$ 2.0 \AA\, it is recorded as a successful prediction.
\begin{myequation}
\text{RMSD} = \sqrt {\frac {\sum_{i=1}^{N} [(x_i - x_i')^2 + (y_i - y_i')^2 + (z_i - z_i')^2]} {N}}
\label{eq:RMSD}
\end{myequation}
where $(x_i, y_i, z_i)$ and $(x_i', y_i', z_i')$ are the Cartesian coordinates of the $i^\text{th}$ atom in two binding poses.
Only non-hydrogen atoms in the molecule were considered in \cite{WorldScientificIJPRAI:Li2014ComparativeAssessmentScoringFunctions}.
Once this analysis was completed over the entire test set, an overall success rate was computed for every scoring function.

\subsection{Evaluation methods of screening power}
As mentioned in CASF-2013, the screening power of a scoring function is defined as the ability of a scoring function to identify the true binders to a given target protein among a pool of random molecules (decoys);
the screening power was evaluated in a cross-docking trial.
The test set includes 65 clusters of PDBbind v2013 core set (195 complexes).
Each cluster consists of 3 complexes formed by a certain protein.
For each protein, the 3 known ligands were taken as positives; whereas the other 195 - 3 = 192 ligands were taken as negatives.
For each of the 65 proteins, all 195 ligands were docked into its binding site, resulting in a total of 65 $\times$ 195 = 12,675 protein-ligand pairs.
Since each protein has 3 different structures, the structure of the best complex in each cluster was selected to be the cluster representative.
Around 50 representative ligand binding poses were selected for each protein-ligand pair.
Each scoring function (in a panel of 20 scoring functions \cite{WorldScientificIJPRAI:Li2014ComparativeAssessmentScoringFunctions}) was applied to score the binding poses of all 195 ligand molecules (including true binders and negatives) of each target protein.
For any given ligand, the best-scored binding pose among all available poses was taken as the predicted binding pose and the corresponding binding score was taken as the predicted binding affinity by this scoring function.
All 195 ligands were then ranked according to their binding scores in a descending order.
The screening power of a scoring function is measured by counting the total number of true binders among the 1\%, 5\%, and 10\% top-ranked ligands.
Enrichment factor (EF) is computed using the following equation \cite{WorldScientificIJPRAI:Li2014ComparativeAssessmentScoringFunctions}:
\begin{myequation}
\text{EF}_\text{x\%} = \frac {\text{NTB}_\text{x\%}} {\text{NTB}_\text{total} \times \text{x}\%}
\end{myequation}
where $\text{NTB}_\text{x\%}$ is the number of true binders observed among the top x\% (where x = 1, 5, or 10) candidates selected by a given scoring function. $\text{NTB}_\text{total}$ is the total number of true binders for the given target protein
(which is typically 3 for each target protein if there are no cross-binders).
As another performance indicator for the screening power in \cite{WorldScientificIJPRAI:Li2014ComparativeAssessmentScoringFunctions};
if the best ligand was found among the 1\%, 5\%, and 10\% top-ranked candidates, a point is counted for the scoring function under test;
an overall success rates on the entire test set at the three different levels are computed accordingly.

\section{Results and Discussion}
\label{sec:ResultsAndDiscussion}









\subsection{Evaluation results of scoring power}

\begin{figure*}[!h]
\centering
\subfloat[RF-Score]
{\includegraphics[scale=0.35]{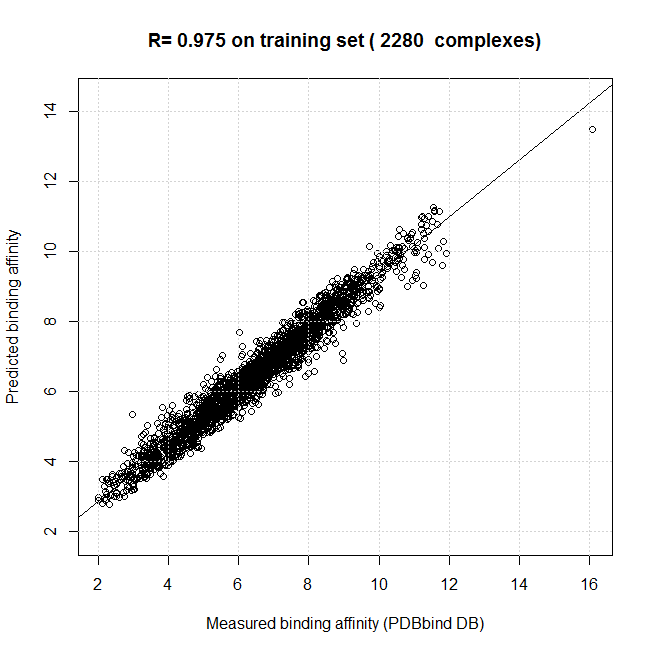}} \qquad
\subfloat[DL-Score]
{\includegraphics[scale=0.35]{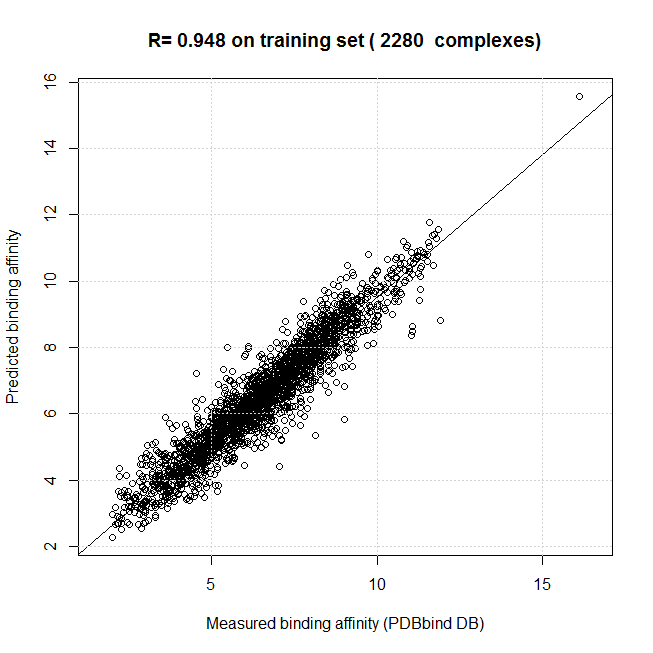}}
\caption{Correlation coefficient of the RF and DL scoring functions on the training set (2280 complexes) on the 108 features.}
\label{fig:LearningTrainingSetAllFeatures}
\end{figure*}

Fig. \ref{fig:LearningTrainingSetAllFeatures} presents the correlation coefficient $R = 0.975$ and $R = 0.948$ of the RF and DL scoring functions respectively on the training set (2280 complexes) on the 108 features.
Despite the superiority of the RF scoring function on the same training set, the DL scoring function achieves better results on the same test set due to using the dropout technique described earlier which avoids over-fitting.

\begin{figure*}[!h]
\centering
\subfloat[RF-Score]
{\includegraphics[scale=0.35]{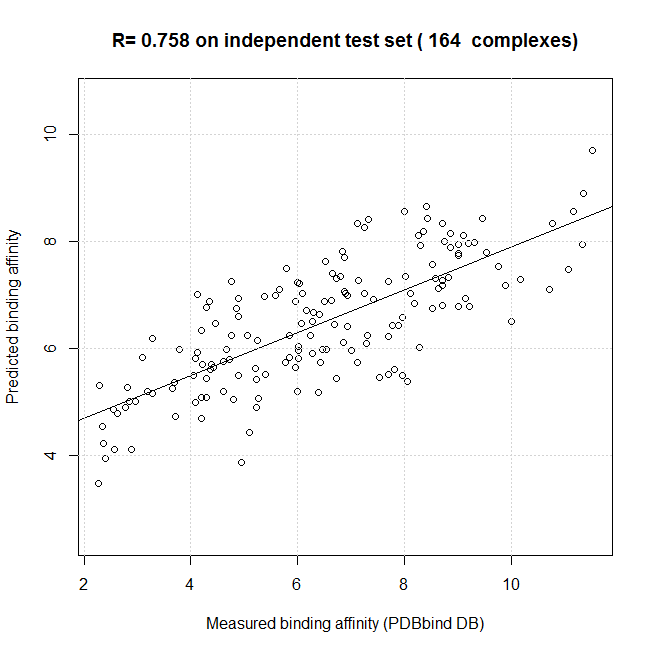}} \qquad
\subfloat[DL-Score]
{\includegraphics[scale=0.35]{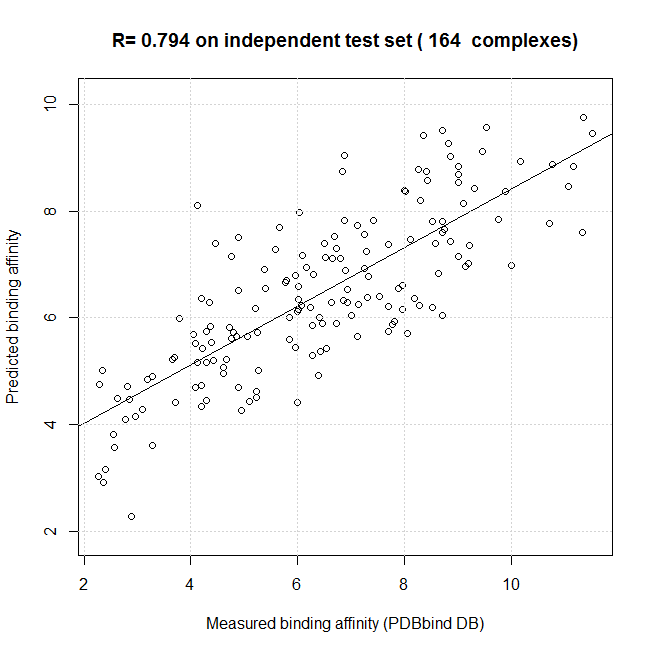}}
\caption{Correlation coefficient of the RF and DL scoring functions on the test set (164 complexes) on the 108 features.}
\label{fig:LearningTestSetAllFeatures}
\end{figure*}

Fig. \ref{fig:LearningTestSetAllFeatures} presents the correlation coefficient $R = 0.758$ and $R = 0.794$ of the RF and DL scoring functions respectively on the independent test set (164 complexes) on the 108 features.
The standard deviation $SD = 1.54$ and $SD = 1.37$ of the RF and DL scoring functions respectively.
Despite the RF scoring function outperforms the best classical scoring function $\text{X-Score}^\text{HM}$, the DL scoring function using the dropout outperforms both scoring functions on the same test set.

The unpaired \emph{t} test results between the RF-Score and DL-Score could be summarized as follows:
(1) the two-tailed P value equals 0.9187,
(2) the mean of RF-Score minus DL-Score equals 0.0154 and the 95\% confidence interval of this difference lies
from -0.2814 to 0.3123, and
(3) the intermediate values used in calculations are: t = 0.1022, df = 326, standard error of difference = 0.151.
Despite by conventional criteria this difference is considered to be not statistically significant,
the Pearson's correlation coefficient has increased by about 5\% of its value (from 0.758 to 0.794) which can be considered an acceptable improvement in this domain.
On the other side, the work presented in \cite{WorldScientificIJPRAI:Wagstaff2012MachineLearningThatMatters} argues that the statistical significance may not be very relevant.
Basically, the interpretation of the results depends on the context, rather than taking mere numbers.
For example, slight increase in performance may be significant in some applications (for example, life critical),
whereas the same increase might be totally irrelevant if the same tool is used in a rather different application.

\begin{figure}[!h]
    \centering
    \includegraphics[width=3.5in, clip,keepaspectratio]{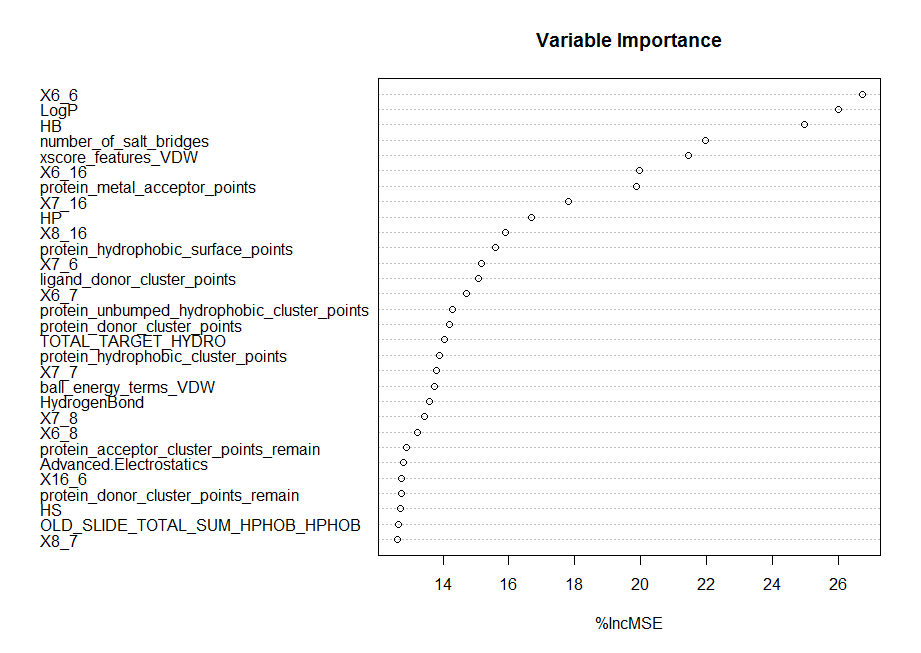}
     \caption{Relative importance of the first 30 features out of the 108 features of the RF scoring function.}
     \label{fig:RandomForestVariableImportanceAllFeatures}
\end{figure}

RF has a built-in tool to measure the importance of individual features across the training set based on the process of ``noising up'' \cite{WorldScientificIJPRAI:Ballester2010PredictingProteinLigandBindingAffinity}.
For each feature, this consists of randomly permuting its values across out-of-bag (OOB) samples \cite{WorldScientificIJPRAI:Svetnik2003RandomForest} for the current tree and evaluating the mean-square-error (MSE) of these perturbed data ($\text{MSE}_j^\text{OOB}$).
The higher the increase in error ($\text{MSE}_{j}^\text{OOB} - \text{MSE}^\text{OOB}$), the more important the $j^{th}$ feature will be for binding affinity prediction \cite{WorldScientificIJPRAI:Ballester2010PredictingProteinLigandBindingAffinity}.
Fig. \ref{fig:RandomForestVariableImportanceAllFeatures} presents the relative importance
of the first 30 features out of the proposed 108 features. Specifically, this figure shows the increase in error observed when individually noising up each of the 108 features; this is an estimate of the importance of the given feature for binding affinity prediction across the training data \cite{WorldScientificIJPRAI:Ballester2010PredictingProteinLigandBindingAffinity}.
Interestingly, the most 30 important features out of the 108 features span all types of features (energy terms, pharmacophore, intermolecular); this highlights the importance of using features of different nature.

\begin{figure}[!h]
    \centering
    \includegraphics[width=3.5in, clip,keepaspectratio]{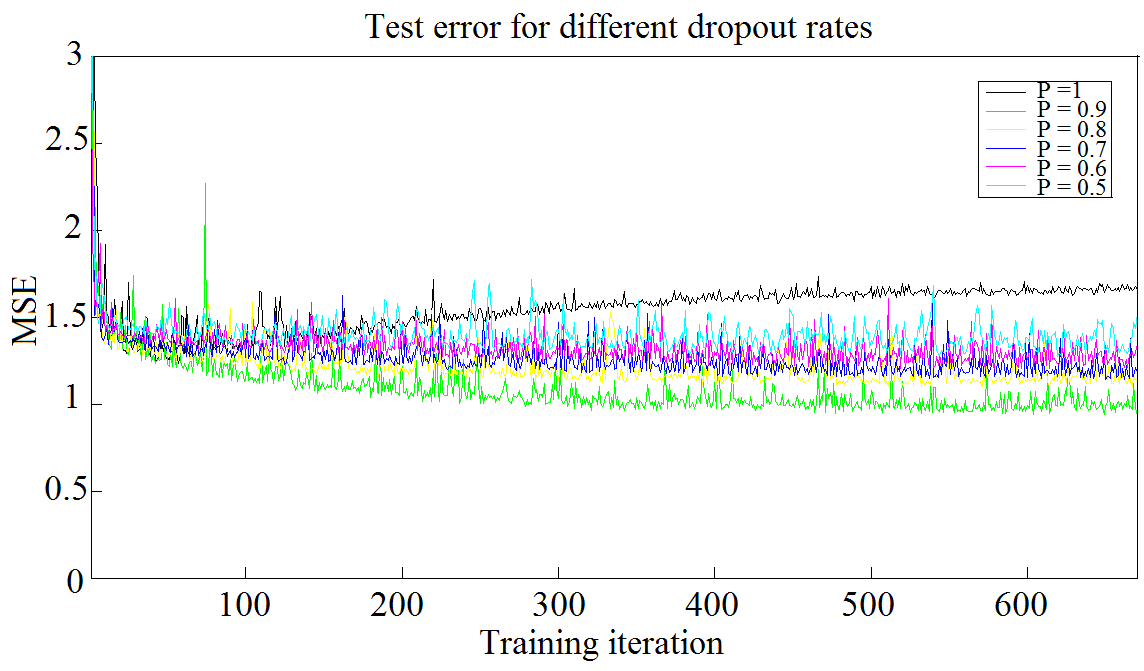}
     \caption{Effect of using the dropout technique on the test set mean square error of the DL scoring function.}
     \label{fig:DeepLearningTestSetErrorDropout}
\end{figure}

We found that the correlation coefficient $R = 0.999$ of the DL scoring function on the training set (2280 complexes) without using dropout in fine tuning (high over-fitting to the training set),
while the correlation coefficient $R = 0.615$ on the independent test set (164 complexes),
i.e., much lower than when using dropout, i.e., $R = 0.794$.
Thus, Fig. \ref{fig:DeepLearningTestSetErrorDropout} presents the effect of using the dropout technique on the MSE between the predicted and experimentally measured binding affinities versus the training iteration (epoch number) of the DL scoring function.
P equals 1 implies no dropout and the optimal performance occurs at P equals 0.9.

We also tried 3 regularization methods (L2, KL-sparsity + L2, and Dropout + L2) and the third one was the best.
While the typical value of the dropout rate for hidden units is in the range from 0.5 to 0.8,
the best value for our model was 0.9 as shown in Fig. \ref{fig:DeepLearningTestSetErrorDropout}.
In addition, using dropout requires to increase the size of the network because dropping units in each iteration reduce its capacity.
Also, no sparsity penalty is used because dropout already forces the hidden units to be sparse.
We tried several architectures with two equally-sized hidden layers, e.g., {[$108 \times 400 \times 400 \times 1$], [$108 \times 500 \times 500 \times 1$], and [$108 \times 600 \times 600 \times 1$]}, and the proposed architecture [$108 \times 500 \times 500 \times 1$] was a compromise between under-fitting of smaller networks and over-fitting of larger networks.

\begin{figure*}[!h]
\centering
\subfloat[DL\_RF-Score]
{\includegraphics[scale=0.35]{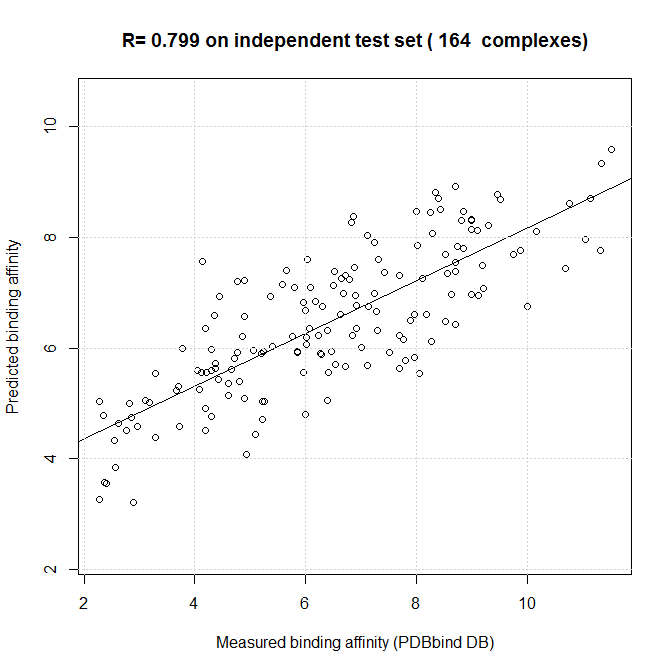}} \qquad
\subfloat[$\text{X-Score}^\text{HM}$]
{\includegraphics[scale=0.36]{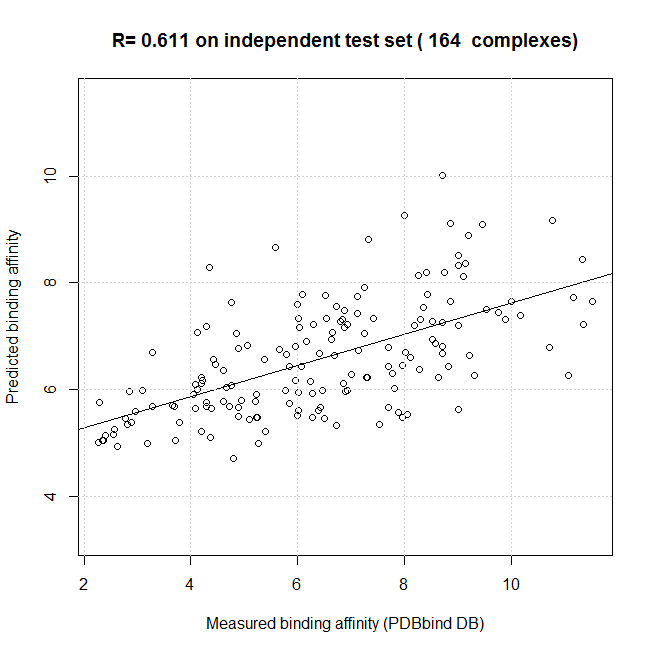}}
\caption{Correlation coefficient of the DL\_RF and $\text{X-Score}^\text{HM}$ scoring functions on the test set (164 complexes) on the 108 features.}
\label{fig:LearningVersusClassicalTestSet}
\end{figure*}

Fig. \ref{fig:LearningVersusClassicalTestSet} presents the correlation coefficient $R = 0.799$ and $R = 0.611$ of the DL\_RF (arithmetic mean = predicted scores sum / 2) and $\text{X-Score}^\text{HM}$ scoring functions respectively on the independent test set (164 complexes).
The standard deviation $SD = 1.41$ and $SD = 1.78$ of the DL\_RF and $\text{X-Score}^\text{HM}$ scoring functions respectively.
Fig. \ref{fig:LearningVersusClassicalTestSet} highlights that combining more than one learning scoring functions can outperform any individual model.
Specifically, these results show the superiority of the learning scoring functions compared to the best classical scoring function $\text{X-Score}^\text{HM}$.

A test of Pearson's correlation significance between the predicted and measured binding affinities of the DL\_RF can be summarized as following: t = 16.8879, df = 162, p-value $<$ 2.2e-16, and
the 95\% confidence interval lies from 0.7353 to 0.8481.

\subsection{Evaluation results of ranking power}


In our experiments, the success rates are calculated based on 50 complete clusters (i.e., with 3 complexes of the same protein) after removing any incomplete clusters in the 164 complexes of the test set.
The DL scoring function ranks the ligands bound to fixed target protein with accuracy 54\% for the high-level
ranking (correctly ranking the 3 ligands bound to the same target protein in a cluster) and with accuracy 78\% for the low-level ranking (correctly ranking the best ligand only in the cluster) while the RF scoring function achieves (46\% and 62\%) respectively whereas the best classical scoring function $\text{X-Score}^\text{HM}$ achieves (54\% and 70\%) respectively.
The Spearman's rank correlation coefficients ($\text{R}_\text{s}$) between the ranks based on the predicted binding affinities and the ranks based on the experimentally measured binding affinities are as following:
DL\_RF-Score = 0.789, DL-Score = 0.787, RF-Score = 0.748, and $\text{X-Score}^\text{HM}$ = 0.627.

\subsection{Impact of increasing the number of training complexes}

In order to perform more validation experiments and in order to study the impact of increasing the number of training complexes on the prediction performance of the learning scoring functions;
we used the whole 2445 = (2281 + 164) complexes of PDBbind v2013;
an increasing training set size (489, 978, 1467, 1956) complexes, and an independent test set of 489 complexes.
Table \ref{tab:ImpactOfIncreasingNumberOfTrainingComplexes} presents the performance of the proposed RF-Score with corresponding $m_\text{best}$ equals (66, 35, 36, 32) respectively versus the DL\_RF-Score where the 4 experiments (489, 978, 1467, 1956) training complexes share the following parameters: momentum= 0.9, dropout rate = 0.9, using 600 epochs, architecture size = [$108 \times 500 \times 500 \times 1$], and batch size = 163.
For the training set size equals (489, 978, 1467, 1956) complexes,
the DL-Score has the following ``learning rate'' and ``L2 weight decay'':
(0.001, 0.01), (0.001, 0.01), (0.001, 0.01), and (0.002, 0.001) respectively.

In Table \ref{tab:ImpactOfIncreasingNumberOfTrainingComplexes}, $R$ is the Pearson's linear correlation coefficient between the predicted and experimentally measured binding affinities,
$R_s$ is the Spearman's rank correlation coefficient between the ranks based on the predicted binding affinities and the ranks based on the measured binding affinities,
RMSE and SD are the root-mean-square-error and standard deviation between the predicted and measured binding affinities respectively.
As it is clear from Table \ref{tab:ImpactOfIncreasingNumberOfTrainingComplexes}, the performance improves by increasing the training set size.
In addition, the performance of the DL\_RF-Score outperforms the performance of the individual RF-Score.

\begin{table}[!h]
\caption{Impact of increasing the number of training complexes on the performance of the learning scoring functions.}
\scriptsize
\begin{tabular}{|l|*{4}{c}|*{4}{c}|}
\hline
$N_\text{train}$  &       \multicolumn{4}{|c|}{RF-Score}       &       \multicolumn{4}{|c|}{DL\_RF-Score}     \\
                  & $R$   & $R_s$ & RMSE &  SD   &  $R$  & $R_s$ & RMSE   & SD  \\
\hline
489               & 0.650 & 0.644 & 1.50  & 1.50 & 0.658 & 0.649 & 1.48   & 1.48 \\
978               & 0.685 & 0.688 & 1.43  & 1.43 & 0.708 & 0.708 & 1.39   & 1.39 \\
1467              & 0.713 & 0.713 & 1.38  & 1.38 & 0.722 & 0.718 & 1.36   & 1.36 \\
1956              & 0.719 & 0.721 & 1.37  & 1.37 & 0.731 & 0.729 & 1.34   & 1.34 \\
\hline
\end{tabular}
\label{tab:ImpactOfIncreasingNumberOfTrainingComplexes}
\end{table}


\subsection{Evaluation results of docking power}

Table \ref{tab:EvaluationResultsDockingPower} presents the performance of the best classical scoring function \cite{WorldScientificIJPRAI:Li2014ComparativeAssessmentScoringFunctions} versus the proposed learning scoring functions in the docking power test when one or more best-scored ligand binding poses are considered.
The cutoff of acceptance is that the RMSD value between one best-scored binding pose and the true binding pose is lower than 2.0 \AA.
The scoring functions are ranked when the top three best-scored ligand poses are considered to match the native pose.

Five complexes were excluded in the docking power test due to the inability of the BALL molecular software for decoy binding poses extraction;
these complexes are of PDB IDs: 3ge7, 3gy4, 1o3f, 1o5b, and 1sqa.
In addition, the complex of PDB ID = 2d1o is also excluded due to incomplete number of extracted decoy binding poses.
Thus, the results are calculated based on the remaining 189 complexes (out of 195 complexes).

As in the scoring and ranking powers test, a DL-based scoring function outperforms the RF scoring function \footnote{A shuffling of the training complexes to match the input to the DL scoring function slightly changes the performance of the RF scoring function in the docking and screening power tests;
this clarifies the slight change in the results here compared to the ones reported in our recent comparative assessment \cite{WorldScientificIJPRAI:Khamis2015ComparativeAssessmentMLScoringFunctionsOnPDBbind2013}.}.
However, the best classical scoring function outperforms the proposed learning scoring functions;
these results comply with the PDBbind v2007 results presented in \cite{WorldScientificIJPRAI:Gabel2014BewareMachineLearningScoringFunctions} and \cite{WorldScientificIJPRAI:Ashtawy2013MolecularDockingDrugDiscoveryMachineLearningApproachesNativePose} (when the binding affinity is used for training).
Nevertheless, the latter work \cite{WorldScientificIJPRAI:Ashtawy2013MolecularDockingDrugDiscoveryMachineLearningApproachesNativePose} proves that the ML scoring functions trained to explicitly predict the RMSD values significantly outperform all classical scoring functions.
In addition, \cite{WorldScientificIJPRAI:Wang2013OptimizationMolecularDockingScores} show that a support vector rank regression (SVRR) algorithm trained with different example datasets, using different training strategies, all achieved increasingly consist accuracies;
in contrast, using the same training datasets, traditional support vector classification and regressions algorithms fail to improve comparably the accuracy of conformation prediction;
the proposed results suggest that with additional features to indicate the comparative fitness between computed binding conformations, the SVRR algorithm has the potential for more accurate docking scores.


\begin{table}[!h]
\caption{Success rates in the docking power test when one or more best-scored ligand binding poses are considered.}
\scriptsize
\centering
\begin{tabular}{|l|*{3}{c|}}
\hline
\textbf{Scoring function}   &   \multicolumn{3}{|c|}{\textbf{Success rates (\%) on}}                                   \\
                            & \textbf{Top pose}         & \textbf{Top two poses}         & \textbf{Top three poses} \\
\hline
ChemPLP@GOLD                & 81.0                          & 86.7                           & 89.7                     \\
DL\_RF                      & 15.9                          & 25.4                           & 36.0                     \\
RF                          & 18.0                          & 24.9                           & 30.2                     \\
DL                          & 13.2                          & 20.6                           & 27.5                     \\
\hline
\end{tabular}
\label{tab:EvaluationResultsDockingPower}
\end{table}

\begin{figure}[!h]
    \centering
    \includegraphics[width=3.5in, clip,keepaspectratio]{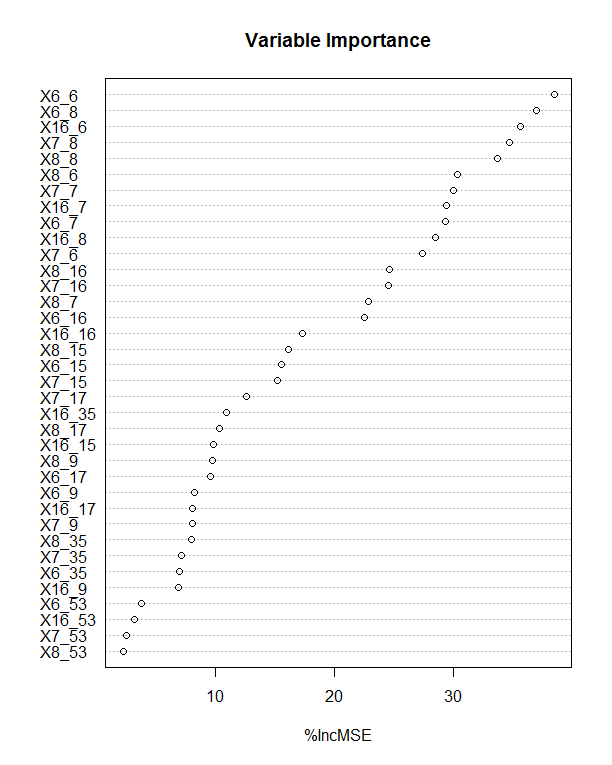}
     \caption{Relative importance of the 36 intermolecular features of the RF scoring function in the docking and screening power tests.}
     \label{fig:RandomForestVariableImportanceAllComplexes}
\end{figure}

Fig. \ref{fig:RandomForestVariableImportanceAllComplexes} presents the relative importance of the 36 intermolecular features of the RF scoring function in the docking and screening power tests.
Among the most important features (\%incMSE $>$ 20), we find the occurrence counts of hydrophobic interactions ($x_{6,6}$), of polar-non-polar contacts ($x_{8,6}, x_{7,6}, x_{6,8}, x_{16,6}$), and also of those intermolecular features correlated with hydrogen bonds ($x_{7,8}, x_{8,8}, x_{8,7}, x_{7,7})$ as reported in \cite{WorldScientificIJPRAI:Ballester2010PredictingProteinLigandBindingAffinity}.

\subsection{Evaluation results of screening power}

Table \ref{tab:EvaluationResultsScreeningPowerEnrichmentFactor} and Table \ref{tab:EvaluationResultsScreeningPowerSuccessRates} present the performance of the best classical scoring function \cite{WorldScientificIJPRAI:Li2014ComparativeAssessmentScoringFunctions} versus the proposed learning scoring functions in the screening power test.
In Table \ref{tab:EvaluationResultsScreeningPowerEnrichmentFactor}, the scoring functions are ranked by their average enrichment factor obtained at the top 1\% level. EF at the top 5\% and top 10\% levels are considerably lower for all scoring functions because the test set in \cite{WorldScientificIJPRAI:Li2014ComparativeAssessmentScoringFunctions} consists of a rather limited number of true binders (normally three) to each target protein.
In Table \ref{tab:EvaluationResultsScreeningPowerSuccessRates}, the scoring functions are ranked by their success rates obtained at the top 1\% level. The numbers in brackets are the number of successful cases, for which the upper limit is 65 (for the learning scoring functions the upper limit is 62); 3 incomplete clusters (only 2 true binders per cluster) were excluded in the screening power test due to the inability of the BALL molecular software for decoy binding poses extraction;
these complexes are of PDB IDs: 3ge7, 1o3f, and 1sqa.
Thus, the results are calculated based on the remaining 62 clusters (out of 65 clusters).

As in the docking power test, a DL-based scoring function outperforms the RF scoring function.
However, the best classical scoring function outperforms the proposed learning scoring functions;
these results comply with the work presented in \cite{WorldScientificIJPRAI:Gabel2014BewareMachineLearningScoringFunctions}, where the authors proposed RF scoring function that do not discriminate DUD-E \cite{WorldScientificIJPRAI:Mysinger2012DUDE} actives from decoys in docking experiments (virtual screening power test).
Thus, the current work together with the work presented in \cite{WorldScientificIJPRAI:Gabel2014BewareMachineLearningScoringFunctions} highlight the need of checking any novel ML scoring function versus the best classical scoring functions with respect to their screening powers.

Some recent research works proposed ML scoring functions which outperform classical scoring functions in the virtual screening test;
this is attributed to the following reason(s):
(1) using SVM target-specific models, e.g., \cite{WorldScientificIJPRAI:Li2011SupportVectorRegression}, \cite{WorldScientificIJPRAI:Kinnings2011MachineLearningImproveDockingScoringFunction}, and
\cite{WorldScientificIJPRAI:Ding2013CharacterizationSmallMoleculeBinding};
this can achieve better performance than training using generic database of different protein families,
specially virtual screening performance is highly dependent on the target receptor being studied as mentioned in \cite{WorldScientificIJPRAI:Durrant2013ComparingNeuralNetworkScoringFunctions},
(2) compensating for potential biases (e.g., chemical properties like small-molecule size and polarizability) leads to improvements in the virtual screening performance \cite{WorldScientificIJPRAI:Durrant2013ComparingNeuralNetworkScoringFunctions},
(3) using DUD-E benchmark \cite{WorldScientificIJPRAI:Mysinger2012DUDE} which includes a total of 102 target proteins as well as 22,886 known ligands for them; while the adopted test set CASF-2013 \cite{WorldScientificIJPRAI:Li2014ComparativeAssessmentScoringFunctions} includes a rather limited number of true binders (normally three) for each target protein, each target protein in DUD-E has several dozens to several hundreds of known binders which is more suitable for ML scoring functions to train on a large number of known binders for each target protein, and
(4) using SVM classifier for discriminating actives from decoys instead of using SVR regressor;
e.g., in order to directly compare the performance of the SVM classifier with that of the SVM regressor in \cite{WorldScientificIJPRAI:Kinnings2011MachineLearningImproveDockingScoringFunction},
the latter was used to rank the DUD InhA data set according to the predicted $\log IC_{50}$ values;
the SVM classifier has been shown to be better than the SVM regressor in virtual screening due to the efficient handling of the data imbalance problem; the data imbalance results in a tendency of the model to classify ligands into the negative class,
and therefore has high precision (due to small false positives) but low recall (due to large false negatives).

\begin{table}[!h]
\caption{Enrichment factors in the screening power test.}
\scriptsize
\centering
\begin{tabular}{|l|*{3}{c|}}
\hline
\textbf{Scoring function}   &     \multicolumn{3}{|c|}{\textbf{Enrichment factor}}                                      \\
                            & \textbf{Top 1\%}              & \textbf{Top 5\%}               & \textbf{Top 10\%}        \\
\hline
GlideScore-SP               & 19.54                         & 6.27                           & 4.14                     \\
DL                          & 2.69                          & 1.40                           & 1.40                     \\
RF                          & 1.61                          & 1.40                           & 1.13                     \\
DL\_RF                      & 1.61                          & 1.29                           & 1.29                     \\
\hline
\end{tabular}
\label{tab:EvaluationResultsScreeningPowerEnrichmentFactor}
\end{table}

\begin{table}[!h]
\caption{Success rates of finding the best ligand molecule in the screening power test.}
\scriptsize
\centering
\begin{tabular}{|l|*{3}{c|}}
\hline
\textbf{Scoring function}   & \multicolumn{3}{|c|}{\textbf{Success rates of finding best ligand molecule among (\%)}}\\
                            & \textbf{Top 1\%}              & \textbf{Top 5\%}               & \textbf{Top 10\%}        \\
\hline
GlideScore-SP               & 60.0 (39)                     & 72.3 (47)                      & 76.9 (50)                \\
DL                          & 6.45 (4)                      & 14.52(9)                       & 25.81(16)                \\
DL\_RF                      & 4.84 (3)                      & 14.52(9)                       & 25.81(16)                \\
RF                          & 4.84 (3)                      & 14.52(9)                       & 24.19(15)                 \\
\hline
\end{tabular}
\label{tab:EvaluationResultsScreeningPowerSuccessRates}
\end{table}

\subsection{Discussion}

In our discussion for the results, we highlight the impact of the dropout technique
that averages a large number of thinned neural networks similarly to model averaging of decision trees used in random forests.
Thus, dropout is crucial in comparison between the two learning scoring functions.
In addition, as mentioned in \cite{WorldScientificIJPRAI:Ashtawy2015BgNScoreBsNScore}, ensemble NN scoring functions are more accurate in predicting the binding affinity of protein-ligand complexes than RF scoring functions;
this is mainly because the ability of neural networks to approximate any underlying function smoothly in contrast to decision trees that model functions with step changes across decision boundaries.
This can interpret why the proposed deep learning scoring function combined with dropout (that is a model averaging technique over a large number of neural networks) outperforms the RF scoring function.
On the other side, for learning more robust features using auto-encoders, noise can be added to the input units;
in \cite{WorldScientificIJPRAI:Vincent2008ExtractingComposingRobustFeaturesDenoisingAutoencoders}, the authors showed that this added noise is helpful in learning deep architectures; dropout also can be interpreted as added noise not only to the input units but also to the hidden units in the neural network and that was very helpful to get more robust features;
this was used in the unsupervised feature learning and in fine-tuning to improve the performance of our network.

Many insights have been learned for both the applications and the methodology (deep learning specifically).
For instance, the unsupervised pre-training phase played important role in training neural networks especially when we increased the number of hidden units, and during parameter optimization we found that some parameters have larger effect on the performance than others, e.g., dropout rate and learning rate. We also found that using KL-sparsity with dropout is not effective and may lead to worse performance.


For the fact that deep learning performs better than random forest in this particular case and its generalization ability;
it is obvious that the efficiency of random forest is mainly due to model averaging of an ensemble of decision trees,
though the dropout technique makes averaging more powerful by averaging large number of thinned networks that allowed the DL scoring function to outperform the RF scoring function.
In \cite{WorldScientificIJPRAI:Srivastava2014Dropout}, the authors show that dropout improved the performance of neural networks on several supervised learning tasks including computational biology; e.g., testing on an alternative splicing dataset to predict the occurrence of alternative splicing based on RNA features.
Thus, in general a deep learning model has many advantages over a random forest model; this is attributed to the deep learning big architecture, unsupervised pre-training, dropout technique, and ability to learn complex functions. Thus, deep learning is expected to outperform random forest in other computational biology prediction tasks.

\begin{figure}[!h]
    \centering
    \includegraphics[scale=0.35, clip,keepaspectratio]{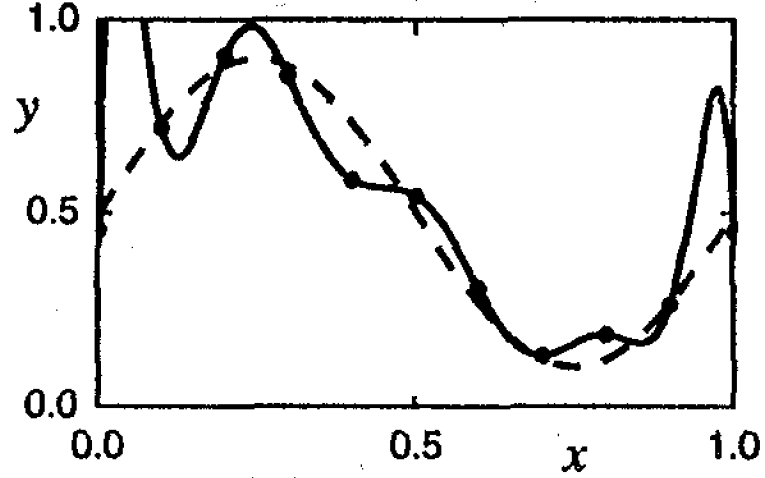}
     \caption{An overfitting example.}
     \label{fig:Overfitting}
\end{figure}

For the fact that deep learning had worse performance for the training set but better performance for the test set;
we have used L2 regularization and dropout to overcome the overfitting on the training dataset;
this results in a higher training error and lower testing error compared to the random forest scoring function;
this suggests that the random forest model fits the noise on the training data more than the deep learning model while the deep learning model achieves better performance on the test set.
For instance, as depicted in Fig. \ref{fig:Overfitting} \cite{WorldScientificIJPRAI:Bishop1995NeuralNetworksPatternRecognition}, the solid curve gives a perfect fit to the training but it gives poor generalization as a result of fitting noise, on the other hand, the dashed curve gives higher training error but lower test error compared to the solid curve.

\section{Conclusion}
\label{sec:Conclusion}




In this paper, we assess the scoring, ranking, docking, and screening powers of deep learning and random forest scoring functions.
For the scoring and ranking powers, the proposed learning scoring functions depend on wide range of features (energy terms, pharmacophore, intermolecular) that entirely characterize the protein-ligand complexes.
For the docking and screening powers, the proposed learning scoring functions depend on the intermolecular features of the RF-Score.

For the scoring power, the DL\_RF scoring function (arithmetic mean between DL and RF scores) achieves Pearson's correlation coefficient between the predicted and experimentally measured binding affinities of 0.799 versus 0.758 of the RF scoring function.
For the ranking power, the DL scoring function ranks the ligands bound to fixed target protein with accuracy 54\% for the high-level ranking and 78\% for the low-level ranking while the RF scoring function achieves (46\% and 62\%) respectively.
For the docking power, the DL\_RF scoring function has a success rate when the three best-scored ligand binding poses are considered within 2 \AA\ root-mean-square deviation from the native pose of 36.0\% versus 30.2\% of the RF scoring function.
For the screening power, the DL scoring function has an average enrichment factor and success rate at the top 1\% level of (2.69 and 6.45\%) respectively versus (1.61 and 4.84\%) respectively of the RF scoring function.

Machine learning scoring functions in general give the ability to utilize as many relevant features as possible (e.g., geometric features and pharmacophore features) and study the impact of those features on the prediction accuracy.
The ensemble-based machine learning approaches (e.g., random forests that are large ensemble of decision trees) are resilient to over-fitting, i.e., give good predictions not only on the training complexes but on any independent test set as well.
Similarly, the deep learning machine learning technique using the dropout approach that is an efficient way to average many neural networks is resilient to over-fitting.
The classical scoring functions in general fail to model the non-linear relationships among individual energy terms.
In addition, the regression coefficients of those individual energy terms are often calibrated based on specific protein family(ies), hence those scoring functions are more prone to over-fitting yielding poor results on independent test sets.
Non-parametric machine learning methods (e.g., random forests) have been successful on generic data sets with multiple classes of target proteins.

As presented by \cite{WorldScientificIJPRAI:Ashtawy2012ComparativeAssessmentRankingAccuracies} and \cite{WorldScientificIJPRAI:Li2014RFCyscore} the performance of the ensemble-based machine learning approaches (e.g., random forest, boosted regression trees) is even much improved either by increasing the number of training complexes or increasing the number of features (opposite to the linear regression-based scoring functions). Thus, as a future work we can study the performance of either the ensemble-based machine learning approaches or the deep learning approach using more training complexes, e.g., from the latest release of the PDBbind benchmark (version 2014) and/or using more features extracted from extra molecular softwares.

\section*{Acknowledgments}

This work is mainly supported by the Information Technology Industry Development Agency (ITIDA) under the ITAC Program Grant no. CFP\#58 and in part by E-JUST Research Fellowship.




%


\bibliographystyle{abbrv}
\bibliography{IEEEabrv,IJPRAIBib}

%
%
%
%
%
%


\vspace*{-0.01in}
\noindent
\rule{12.6cm}{.1mm}

%
%
%

\end{document}